\documentclass[aps,prl,showpacs,twocolumn]{revtex4}
\usepackage{epsfig}
\begin{document}

\title
{A {\it universal} programmable quantum state discriminator that is optimal 
for unambiguously distinguishing between {\it unknown} states} 
\author{J\'{a}nos A. Bergou}
\author{Mark Hillery}
\affiliation{Department of Physics, Hunter College, City
    University of New York, 695 Park Avenue, New York, NY 10021,
    USA}

\date{\today}
\begin{abstract}
We construct a device that can unambiguously discriminate between two unknown 
quantum states. The unknown states are provided as inputs, or  programs, for 
the program registers and a third system, which is guaranteed to be prepared 
in one of the states stored in the program registers, is fed into the data 
register of the device.  The device will then, with some probability of 
success, tell us whether the unknown state in the data register matches the 
state stored in the first or the second program register. We show that the 
optimal device, i. e. the one that maximizes the probability of success, is 
universal. It does not depend on the actual unknown states that we wish to 
discriminate.
\end{abstract}


\maketitle

Given two unknown quantum states, $|\psi_{1}\rangle$ 
and $|\psi_{2}\rangle$, we can construct a device that will  unambiguously 
discriminate between them.  If this device is given a system in one of these 
two states, it will produce one of three outputs, 1, 2, or 0.  If the output 
is 1, then the input was $|\psi_{1}\rangle$, if the output is 2, then the 
input was $|\psi_{2}\rangle$, and if the output is 0, which we call failure, 
then we learn nothing about the input.  The device will not make an error, it 
will never give an output of 2 if the input was $|\psi_{1}\rangle$, and vice 
versa.  This strategy is called unambiguous discrimination.  The input 
states are not necessarily orthogonal; in fact, they can be completely 
arbitrary within the constraint that they are linearly independent 
\cite{chefles}.  The cost 
associated with this condition is that the probability of receiving the output
0 (failure) is not zero.  The minimum value of this probability for two known 
and equally likely states is 
$|\langle\psi_{1}|\psi_{2}\rangle|$ (Refs. \cite{ivanovic}-\cite{peres}).  

The actual state-distinguishing device for two {\it known} states depends on 
the two states, $|\psi_{1}\rangle$ and $|\psi_{2}\rangle$, i. e. these two 
states are ``hard wired'' into the machine.  What we would like to do here is 
to see if we can construct a machine in which the information about 
$|\psi_{1}\rangle$ and $|\psi_{2}\rangle$ is supplied in the form of a 
program.  This machine would be capable, with the correct program, of 
distinguishing any two quantum states.  One such device has been proposed by 
Du\v{s}ek and Bu\v{z}ek \cite{dusek}.  This device distinguishes the two 
states $\cos (\phi /2)|0\rangle \pm \sin (\phi /2)|1\rangle$.  
The angle $\phi$ is encoded into 
a one-qubit program state in a somewhat complicated way.  The performance of 
this device is good; it does not achieve the maximum possible success 
probability for all input states, but its success 
probability, averaged over the angle $\phi$, is greater than 90\% of the 
optimal value.  In a series of recent works Fiur\'{a}\v{s}ek \emph{et al.} 
investigated a closely related programmable device that can perform a von 
Neumann projective measurement in any basis, the basis being specified by
the program.  Both deterministic and probabilistic approaches were explored 
\cite{fiurasek}, and experimental versions of both the state discriminator 
and the projective measurement device were realized \cite{soubusta}.  Sasaki
\emph{et al.} developed a related device, which they called a quantum 
matching machine \cite{sasaki}.  Its input consists of $K$ copies of two
equatorial qubit states, which are called templates, and $N$ copies of
another equatorial qubit state $|f\rangle$.  The device determines to which
of the two template states $|f\rangle$ is closest.  This device does not
employ the unambiguous discrimination strategy, but optimizes an average
score that is related to the fidelity of the template states and $|f\rangle$. 
Programmable quantum devices to accomplish other tasks have recently been 
explored by a number of authors \cite{nielsen}-\cite{paz}.

Here we want to construct a programmable state discriminating machine
whose program is related in a simple way to the states $|\psi_{1}\rangle$ 
and $|\psi_{2}\rangle$ that we are trying to distinguish.  A motivation 
for our problem is that the program state may be 
the result of a previous set of operations in a quantum information processing
device, and it would be easier to produce a state in which the information 
about $|\psi_{1}\rangle$ and $|\psi_{2}\rangle$ is encoded in a simple way 
that one in which the encoding is more complicated.

We shall, therefore, consider the following problem which is perhaps the 
simplest version of a programmable state discriminator.  The program consists 
of the two qubit states that we wish to distinguish.  In other words, we are 
given two qubits, one in the state $|\psi_{1}\rangle$ and another in the state
$|\psi_{2} \rangle$.  We have no knowledge of the states $|\psi_{1}\rangle$ 
and $|\psi_{2}\rangle$.  Then we are given a third qubit that is guaranteed to
be in one of these two program states, and our task is to determine, as best 
we can, in which one.  We are allowed to fail, but not to make a mistake.  
What is the best procedure to accomplish this?

We shall consider the first two qubits we are 
given as a program.  They are fed into the program register of some device, 
called the programmable state discriminator, and the third, unknown qubit is 
fed into the data register of this device.  The device then tells us, with 
optimal probability of success, which one of the two program states the 
unknown state of the qubit in the data register corresponds to.  We can
design such a device by viewing our problem as a task in measurement 
optimization.  We want to find
a measurement strategy that, with maximal probability of success, 
will tell us which one of the two program states, stored in the program 
register, matches the unknown  state, stored in the data register.  Our 
measurement is allowed to return an inconclusive result but never an erroneous
one.  Thus, it will be described by a POVM (positive-operator-valued
measure) that will return 1 (the unknown state stored in the data register 
matches $|\psi_{1}\rangle$), 2 (the unknown state stored in the data register 
matches $|\psi_{2}\rangle$), or 0 (we do not learn anything about the 
unknown state stored in the data register).

Our task is then reduced to the following measurement optimization problem.
One has two input states
\begin{eqnarray}
|\Psi_{1}^{in}\rangle & = & |\psi_{1}\rangle_{A}|\psi_{2}\rangle_{B}
|\psi_{1}\rangle_{C} \ , \nonumber \\
|\Psi_{2}^{in}\rangle & = & |\psi_{1}\rangle_{A}|\psi_{2}\rangle_{B}
|\psi_{2}\rangle_{C} ,
\end{eqnarray}
where the subscripts A and B refer to the program registers (A contains 
$|\psi_{1}\rangle$ and B contains $|\psi_{2}\rangle$), and the subscript C 
refers to the data register.  Our goal is to unambiguously distinguish between
these inputs, keeping in mind that one has no knowledge of $|\psi_{1}\rangle$ 
and $|\psi_{2}\rangle$, i. e. we want to find a POVM that will 
accomplish this.

Let the elements of our POVM be $\Pi_{1}$, corresponding to unambiguously 
detecting $|\Psi_{1}^{in}\rangle$, $\Pi_{2}$, corresponding to unambiguously 
detecting $|\Psi_{2}^{in}\rangle$, and $\Pi_{0}$, corresponding to failure.   
The probabilities of successfully identifying the two possible 
input states are given by
\begin{equation}
\langle\Psi_{1}^{in}|\Pi_{1}|\Psi_{1}^{in}\rangle = p_{1} \hspace{1cm} 
\langle\Psi_{2}^{in}|\Pi_{2}|\Psi_{2}^{in}\rangle = p_{2} \ ,
\label{probs}
\end{equation}
and the condition of no errors implies that
\begin{equation}
\Pi_{2}|\Psi^{in}_{1}\rangle = 0 \hspace{1cm} 
\Pi_{1}|\Psi^{in}_{2}\rangle = 0 \ .
\label{UDcond} 
\end{equation}
In addition, because the alternatives represented by the POVM exhaust all
possibilities, we have that 
\begin{equation}
I = \Pi_{1} + \Pi_{2} + \Pi_{0} \ .
\end{equation}

The fact that we know nothing about $|\psi_{1}\rangle$ and $|\psi_{2}
\rangle$ means that the only way we can guarantee satisfying the above
conditions is to take advantage of the symmetry properties of the states,
i.e.\ that $|\Psi_{1}^{in}\rangle$ is invariant under interchange of the 
first and third qubits, and $|\Psi_{2}^{in}\rangle$ is invariant under
interchange of the second and third qubits.  That unknown 
states can be unambiguously compared with a non-zero probability of success, 
using symmetry considerations only, has been first pointed out by Barnett 
{\it et al.} \cite{BCJ}.  In our case, we require that $\Pi_{1}$ 
give zero when acting on states that are symmetric in qubits $B$ 
and $C$, while $\Pi_{2}$  give zero when acting on states that are
symmetric in qubits $A$ and $C$.  Defining the antisymmetric states for the 
corresponding pairs of qubits
\begin{eqnarray}
|\psi ^{(-)}_{BC}\rangle & = &\frac{1}{\sqrt{2}}(|0\rangle_{B}
|1\rangle_{C}-|1\rangle_{B}|0\rangle_{C}) \ ,  \nonumber \\
|\psi ^{(-)}_{AC}\rangle & = &\frac{1}{\sqrt{2}}(|0\rangle_{A}
|1\rangle_{C}-|1\rangle_{A}|0\rangle_{C}) \ ,
\end{eqnarray}
we introduce the projectors to the antisymmetric subspaces of the 
corresponding qubits as
\begin{eqnarray}
P_{BC}^{as}&=&|\psi ^{(-)}_{BC}\rangle \langle \psi ^{(-)}_{BC}|  \ ,  
\nonumber \\
P_{AC}^{as}&=&|\psi ^{(-)}_{AC}\rangle \langle \psi ^{(-)}_{AC}|  \ .
\end{eqnarray}
We can now take for $\Pi_{1}$ and $\Pi_{2}$ the operators
\begin{eqnarray}
\Pi_{1} & = & c_{1} I_{A}\otimes P_{BC}^{as} \ , \nonumber \\
\Pi_{2} & = & c_{2} I_{B}\otimes P_{AC}^{as} \ ,
\label{SuccesOps1}
\end{eqnarray}
where $I_{A}$ and $I_{B}$ are the identity operators on the spaces of qubits 
$A$ and $B$, respectively, and $c_{1}$ and $c_{2}$ are as yet undetermined 
nonnegative real numbers.  The no-error condition dictates that $\Pi_{1}=
Q_{A}\otimes P_{BC}^{as}$ and $\Pi_{2}=Q_{B}\otimes P_{AC}^{as}$, and it
can be shown that the unknown operators $Q_{A}$ and $Q_{B}$ can be chosen to
be proportional to the identity \cite{bergou}.
Using the above expressions for $\Pi_{j}$, where $j=1,2$ in Eq. (\ref{probs}),
we find that 
\begin{eqnarray}
p_{j} = \langle\Psi_{j}^{in}|\Pi_{j}|\Psi_{j}^{in}\rangle & = & 
c_{j} \frac{1}{2}(1-|\langle\psi_{1}|\psi_{2}\rangle |^{2}) \ .
\label{pwithc}
\end{eqnarray}

The average probability, $P$, of successfully determining which state we have,
assuming that the input states occur with a probability of $\eta_{1}$ and 
$\eta_{2}$, respectively, is given by
\begin{equation}
P = \eta_{1} p_{1} + \eta_{2} p_{2} = \frac{1}{2}(\eta_{1} c_{1} 
+ \eta_{2} c_{2}) (1-|\langle\psi_{1}|\psi_{2}\rangle|^{2})  \ ,
\label{Psuccess}
\end{equation}
and we want to maximize this expression subject to the constraint
that $\Pi_{0} = I-\Pi_{1} - \Pi_{2}$ is a positive operator.

Let $S$ be the 4-dimensional subspace of the entire eight-dimensional Hilbert 
space of the three qubits, A, B, and C, that is spanned by the vectors 
$|0\rangle_{A}|\psi^{(-)}_{BC}\rangle$, $|1\rangle_{A}|\psi^{(-)}_{BC}
\rangle$, $|0\rangle_{B}|\psi^{(-)}_{AC)}\rangle$, and $|1\rangle_{B}|
\psi^{(-)}_{AC}\rangle$.  In the orthogonal complement of $S$, $S^{\perp}$, 
the operator $\Pi_{0}$ acts as the identity, so that in $S^{\perp}$, 
$\Pi_{0}$ is positive.  Therefore, we need to investigate its action in $S$.  
First, let us construct an orthonormal basis for $S$.  Applying the 
Gram-Schmidt process to the four vectors, given above, that span $S$, we 
obtain the orthonormal basis
\begin{eqnarray}
|\Phi_{1}\rangle & = & |0\rangle_{A}|\psi^{(-)}_{BC}\rangle \ , \nonumber \\
|\Phi_{2}\rangle & = & \frac{1}{\sqrt{3}}(2 |0\rangle_{B}|\psi^{(-)}_{AC}
\rangle - |0\rangle_{A}|\psi^{(-)}_{BC}\rangle )  \ , \nonumber \\
|\Phi_{3}\rangle & = & |1\rangle_{A}|\psi^{(-)}_{BC}\rangle \ ,\nonumber \\
|\Phi_{4}\rangle & = & \frac{1}{\sqrt{3}}(2|1\rangle_{B}|\psi^{(-)}_{AC}
\rangle - |1\rangle_{A}|\psi^{(-)}_{BC}\rangle ) .
\end{eqnarray}
In this basis, the operator $\Pi_{0}$, restricted to the subspace $S$, is
given by the $4\times 4$ matrix
\begin{widetext}
\begin{equation}
\Pi_{0} = \left(\begin{array}{cccc} 1- c_{1}-\frac{1}{4}c_{2} &
-\frac{\sqrt{3}}{4}c_{2} & 0 & 0 \\
-\frac{\sqrt{3}}{4}c_{2} & 1 - \frac{3}{4}c_{2} & 0 & 0 \\
0 & 0 & 1- c_{1}-\frac{1}{4}c_{2} & -\frac{\sqrt{3}}{4}c_{2} \\
0 & 0 & -\frac{\sqrt{3}}{4}c_{2} & 1 - \frac{3}{4}c_{2} 
\end{array}\right)
\end{equation}
\end{widetext}
Because of the block diagonal nature of $\Pi_{0}$, the characteristic equation
for its eigenvalues, $\lambda$, is given by the biquadratic equation
\begin{equation}
[\lambda^{2} - (2-c_{1} - c_{2}) \lambda + 1 - c_{1}
- c_{2} + \frac{3}{4}c_{1}c_{2}]^{2} = 0 \ .
\end{equation}

It is easy to obtain the eigenvalues explicitly. For our purposes, however, 
the conditions for their nonnegativity are more useful. These can be read out 
from the above equation, yielding
\begin{eqnarray}
2-c_{1} - c_{2} & \geq & 0 \ , \nonumber \\
1 - c_{1} - c_{2} +\frac{3}{4}c_{1}c_{2} & \geq & 0 \ .
\label{cond}
\end{eqnarray}
The second is the stronger of the two conditions. When it is satisfied the 
first one is always met but the first one can still be used to eliminate 
nonphysical solutions. We can use the second condition to express 
$c_{2}$ in terms of $c_{1}$,
\begin{equation}
c_{2} \leq \frac{1-c_{1}}{1-(3/4)c_{1}} \ .
\label{cond2}
\end{equation}
For maximum probability of success, we chose the equal sign. Inserting the 
resulting expression into (\ref{Psuccess}) gives 
\begin{equation}
P = \frac{1}{2}(\eta_{1} c_{1} + \eta_{2} \frac{1-1c_{1}}{1-(3/4)c_{1}})
(1-|\langle\psi_{1}|\psi_{2}\rangle|^{2}) \ .
\end{equation}
We can easily find $c_{1}=c_{1,opt}$ where the right-hand side 
of this expression is maximum and using this together with Eq. (\ref{cond2}) 
we obtain
\begin{equation}
c_{1,opt} = \frac{2}{3}\left( 2 - \sqrt{\frac{\eta_{2}}{\eta_{1}}}
\right)  \hspace{0.3cm} c_{2,opt} = \frac{2}{3}\left( 2 - \sqrt{\frac
{\eta_{1}}{\eta_{2}}}\right) .
\label{p12opt}
\end{equation} 
Inserting these optimal values into (\ref{Psuccess}) gives
\begin{equation}
P_{POVM} = \frac{2}{3}(1-\sqrt{\eta_{1} \eta_{2}}) (1-|\langle\psi_{1}|
\psi_{2}\rangle|^{2}) \ .
\label{Ppovm}
\end{equation}

This is not the full story, however. The above expression is valid only when 
$c_{1,opt}$ and $c_{2,opt}$ are both non-negative.  From Eq. 
(\ref{p12opt}) it is easy to see that this holds if
\begin{equation}
\frac{1}{5} \leq \eta_{1} , \eta_{2} \leq \frac{4}{5} \ .
\label{boundaries}
\end{equation}
In order to understand what happens outside this interval, we have to turn our
attention to the detection operators.  Using $c_{1,opt}$ and 
$c_{2,opt}$ in Eq. (\ref{SuccesOps1}) yields
\begin{eqnarray}
\Pi_{1,opt} & = & \frac{2}{3}\left( 2 - \sqrt{\frac{\eta_{2}}{\eta_{1}}}
\right)   I_{A}\otimes P^{as}_{BC} \ , \nonumber \\
\Pi_{2,opt} & = & \frac{2}{3}\left( 2 - \sqrt{\frac{\eta_{1}}{\eta_{2}}}
\right)  I_{B}\otimes P^{as}_{AC} \ .
\label{POVMelements2}
\end{eqnarray}
For $\eta_{1}=\frac{4}{5}$ (and $\eta_{2}=\frac{1}{5}$), $\Pi_{1,opt}=I_{A} 
P^{as}_{BC}$ and $\Pi_{2,opt}=0$. This structure then remains valid for 
$\eta_{1} \geq \frac{4}{5}$. In other words, when the first input dominates 
the preparation it is advantageous to use the full projector that 
distinguishes it with maximal probability of success, $p_{1,opt}
= \frac{1}{2} (1-|\langle\psi_{1}|\psi_{2}\rangle|^{2})$, at the expense of sacrificing the second input completely, 
$p_{2,opt} = 0$. These values yield the average success probability,
\begin{equation}
P_{1} = \frac{1}{2}\eta_{1}(1-|\langle\psi_{1}|\psi_{2}\rangle|^{2}) \ ,
\label{P1}
\end{equation}
for $\eta_{1} \geq \frac{4}{5}$.  Conversely, for $\eta_{2}=\frac{4}{5}$, 
$\Pi_{2,opt}=I_{B} P^{as}_{AC}$ and $\Pi_{1,opt}=0$. This structure then 
remains valid for $\eta_{2} \geq \frac{4}{5}$. So, when the second input 
dominates the preparation it is advantageous to use the full projector that 
distinguishes it with maximal probability of success, $p_{2,opt}= \frac{1}{2} (1-|\langle\psi_{1}|\psi_{2}\rangle|^{2})$,
at the expense of sacrificing the first input completely, 
$p_{1,opt}=0$.  These values yield the average success probability,
\begin{equation}
P_{2} = \frac{1}{2}\eta_{2}(1-|\langle\psi_{1}|\psi_{2}\rangle|^{2}) \ ,
\label{P2}
\end{equation}
for $\eta_{2} \geq \frac{4}{5}$.  As we see, the situation is fully symmetric 
in the inputs and {\it a priori} probabilities.  In the intermediate range, 
neither one of the inputs dominates the preparation, and we want to identify 
them as best as we can, so the POVM solution will do the job there.
Our findings can be summarized as follows
\begin{eqnarray}
    \label{Popt}
    P^{opt} = \left\{ \begin{array}{ll}
    P_{POVM} & \mbox{ if
    $\frac{1}{5} \leq \eta_{1}
    \leq  \frac{4}{5}$} \ , \\
    P_{2} & \mbox{ if $\eta_{1} <
    \frac{1}{5}$} \ , \\ 
    P_{1} & \mbox{ if $\frac{4}{5}
    < \eta_{1}$} \ .
    \end{array}
    \right.
\end{eqnarray}

Equation (\ref{Popt}) represents our main result. In the 
intermediate range of the {\em a priori} probability the optimal failure 
probability, Eq. (\ref{Ppovm}), is achieved by a generalized measurement 
or POVM. Outside this region, for very small {\em a priori} probability, 
$\eta_{1} \leq 1/5$, when the preparation is dominated by the second input, 
or very large {\em a priori} probability, $\eta_{1} \geq 4/5$, when the 
preparation is dominated by the first input, the optimal failure 
probabilities, Eqs. (\ref{P1}) and (\ref{P2}), are realized by standard von 
Neumann measurements. For very small $\eta_{1}$ the optimal von Neumann 
measurement is a projection onto the antisymmetric subspace of the A and C 
qubits.  For very large $\eta_{1}$ the optimal von Neumann measurement is a 
projection onto the antisymmetric subspace of the B and C qubits.  At the 
boundaries of their respective regions of validity, the optimal measurements 
transform into one another continuously.  We also see that the results depend 
on the overlap of the unknown states only.  If we do not know the states but 
we know their overlap then Eqs. (\ref{Ppovm}), (\ref{P1}), and (\ref{P2}) 
immediately give the optimal solutions for this situation. If we know nothing 
about the states, not even their overlap, then we average these expressions 
over all input states, which results in the 
factor, $1-|\langle\psi_{1}|\psi_{2}\rangle|^{2}$, being replaced by its 
average value of $\frac{1}{2}$. Then we have the optimum average probabilities
of success in the various regions. This situation is depicted in Fiq. 1.

\begin{figure}[ht]
\epsfig{file=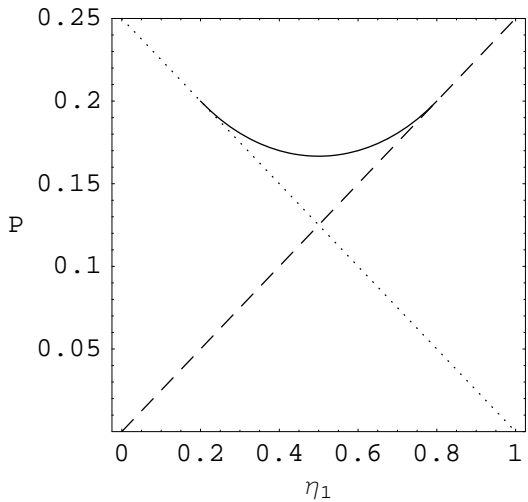, height=7cm}
\caption{Optimal average success probability, $P$, vs. the {\it a priori} 
probability, $\eta_{1}$. Dashed line: $P_{1}$ from Eq. (\ref{P1}), dotted 
line: $P_{2}$ from Eq. (\ref{P2}), and solid line: $P_{POVM}$ from Eq. 
(\ref{Ppovm}). For the figure we replaced $1-|\langle\psi_{1}|\psi_{2}
\rangle|^{2}$ by its average, $\frac{1}{2}$.  The optimal $P$ is given by 
$P_{2}$ for $\eta_{1}<0.2$, by $P_{POVM}$ for $0.2 \leq \eta_{1} \leq
0.8$ and by $P_{1}$ for $0.8<\eta_{1}$.}
\label{Fig1}
\end{figure}

In its range of validity the POVM performs better than any von Neumann 
measurement that does not introduce errors. 
From the figure it also can be read out that the difference 
between the performance of the POVM and that of the von Neumann projective 
measurements is largest for $\eta_{1}=\eta_{2}=\frac{1}{2}$. For these values
$P_{POVM}^{ave}=\frac{1}{6}$ while $P_{1}^{ave}=P_{2}^{ave}=\frac{1}{8}$ so 
the POVM represents a 33\% improvement over the standard quantum measurement.

Finally, we want to point out a striking feature of the programmable state 
discriminator. Neither the optimal detection operators nor the boundaries for 
their region of validity, Eqs. (\ref{boundaries}) and (\ref{POVMelements2}), 
depend on the unknown states. Therefore, our device is {\it universal}, it 
will perform optimally for any pair of unknown states. Only the probability of
success for fixed but unknown states will depend on the overlap of the states. 

This POVM, then, provides us with the best procedure for solving the
problem posed at the beginning of this paper. It also demonstrates the
role played by \emph{a priori} information.  This device has a smaller
success probability than one designed for a case in which we know one of
the input states \cite{bergou}, which in turn has a smaller success
probability than one designed for the case when we know both possible
input states.  While its success probability is lower than that for a device that distinguishes known states, the device discussed here is more flexible. All of
the information about the states is carried by a quantum program, which
means that it works for any two states.  Consequently, it can be used as 
part of a larger device that produces quantum states that need to be
unambiguously identified.

\begin{acknowledgments}
This research was partially supported by a grant from the Humboldt Foundation (JB) and the National Science Foundation under 
grant PHY 0139692 (MH). JB also acknowledges helpful discussions with Prof. W. Schleich and his group during a visit to the University of Ulm. \end{acknowledgments}

\bibliographystyle{unsrt}

\begin{thebibliography}{99}
\bibitem{chefles} A.\ Chefles, Phys.\ Lett.\ A \textbf{239}, 339  (1998).
\bibitem{ivanovic} I.\ D.\ Ivanovic, Phys.\ Lett.\ A {\bf 123}, 257 (1987).
\bibitem{dieks} D.\ Dieks, Phys.\ Lett.\ A {\bf 126}, 303 (1988).
\bibitem{peres} A.\ Peres, Phys.\ Lett.\ A {\bf 128}, 19 (1988).
\bibitem{dusek} M.\ Du\v{s}ek and V.\ Bu\v{z}ek, \pra {\bf 66}, 022112 (2002).
\bibitem{fiurasek}J.\ Fiur\'{a}\v{s}ek, M.\ Du\v{s}ek, and R.\ Filip, Phys.\ 
Rev.\ Lett.\ {\bf 89}, 190401 (2002); J.\ Fiur\'{a}\v{s}ek and M.\ Du\v{s}ek,
Phys.\ Rev.\ A {69}, 032302 (2004).
\bibitem{soubusta}J.\ Soubusta, A.\ \v{C}ernoch, J.\ Fiur\'{a}\v{s}ek, and 
M.\ Du\v{s}ek,\ Phys.\ Rev.\ A {\bf 69}, 052321.
\bibitem{sasaki}M.\ Sasaki and A.\ Carlini, Phys.\ Rev.\ A {\bf 66},
022303 (2002); M.\ Sasaki, A.\ Carlini, and R.\ Jozsa, Phys.\ Rev.\ A
{\bf 64}, 022317 (2001).
\bibitem{nielsen} M.\ Nielsen and I.\ Chuang, Phys.\ Rev.\ Lett.\ {\bf 79},
321 (1997).
\bibitem{preskill} J.\ Preskill, Proc. Roy.\ Soc.\ Lond.\ A {\bf 454}, 385
(1998).
\bibitem{hillery} M.\ Hillery, V.\ Bu\v{z}ek, and M.\ Ziman, \pra {\bf 65}, 
022301 (2002).
\bibitem{vidal}G.\ Vidal, L.\ Masanes, and I.\ Cirac, Phys.\ Rev.\ Lett.\ 
{\bf 88}, 047905 (2002).
\bibitem{hillery2} M.\ Hillery, V.\ Bu\v{z}ek, and M.\ Ziman, \pra {\bf 66}, 
032302 (2002). 
\bibitem{ekert} A.\ K.\ Ekert, C.\ M.\ Alves, D.\ K.\ L.\ Oi, M.\ Horodecki,
P.\ Horodecki, and L.\ C.\ Kwek, Phys.\ Rev.\ Lett.\ {\bf 88}, 217901
(2002).
\bibitem{paz}J.\ Paz and A.\ Roncaglia, quant-ph/0306143.
\bibitem{BCJ} S.\ M.\ Barnett, A.\ Chefles, and I.\ Jex, Phys.\ Lett.\ A 
{\bf 307}, 189 (2003). 
\bibitem{bergou} J.\ Bergou, M.\ Hillery, and V.\ Bu\v{z}ek, in preparation.
\end{thebibliography}

\end{document}